\documentclass[useAMS,usenatbib]{mn2e}

\usepackage{times}
\usepackage{mathptmx}
\usepackage{graphicx}

\usepackage{graphicx}
\usepackage{color}

\title[Cuspy triaxial stellar systems]{Models of cuspy triaxial stellar systems. III:
The effect of velocity anisotropy on chaoticity} 
\author[D. D. Carpintero, J. C. Muzzio and H. D. Navone]
{D. D. Carpintero$^{1,2}$,
\thanks{E-mail: ddc@fcaglp.unlp.edu.ar}
J. C. Muzzio$^{1,2}$,
\thanks{E-mail: jcmuzzio@fcaglp.unlp.edu.ar}
and H. D. Navone$^{3,4}$\\
$^{1}$Facultad de Ciencias Astron\'omicas y Geof\'{i}sicas,
Universidad Nacional de La Plata, La Plata, Argentina\\
$^{2}$Instituto de Astrof\'{i}sica de La Plata (CONICET La Plata--UNLP) \\
$^{3}$Facultad de Ciencias Exactas, Ingenier\'{i}a y Agrimensura,
Universidad Nacional de Rosario, Rosario, Argentina\\
$^{4}$Instituto de F\'\i sica Rosario (CONICET--UNR)}

\begin{document}

\date{}

\pagerange{\pageref{firstpage}--\pageref{lastpage}} \pubyear{}

\maketitle

\label{firstpage}

\begin{abstract}

In several previous investigations we presented models of triaxial stellar
systems, both cuspy and non cuspy, that were highly stable and harboured large
fractions of chaotic orbits. All our models had been obtained through cold
collapses of initially spherical $N$--body systems, a method that necessarily
results in models with strongly radial velocity distributions. Here we
investigate a different method that was reported to yield cuspy triaxial models
with virtually no chaos. We show that such result was probably due to the use of
an inadequate chaos detection technique and that, in fact, models with
significant fractions of chaotic orbits result also from that method. Besides,
starting with one of the models from the first paper in this series, we obtained
three different models by rendering its velocity distribution much less radially
biased (i.e., more isotropic) and by modifying its axial ratios through
adiabatic compression. All three models yielded much higher fractions of regular
orbits than most of those from our previous work. We conclude that it is
possible to obtain stable cuspy triaxial models of stellar systems whose
velocity distribution is more isotropic than that of the models obtained from
cold collapses. Those models still harbour large fractions of chaotic orbits
and, although it is difficult to compare the results from different models, we
can tentatively conclude that chaoticity is reduced by velocity isotropy.

\end{abstract}


\section{Introduction}

A simple way to build stable models of triaxial stellar systems is to start from
a spherical distribution of point masses and to follow its collapse with an
$N$--body code. If the initial velocity dispersion is low (i.e., if the original
distribution is $\it{cold}$) the radial orbit instability leads the evolution of
the system towards a triaxial stable state \citep{AM90}. Models that are cuspy
(i.e., where near the centre the density, $\rho(r)$, is proportional to
$r^{-\gamma}$, where $r$ is the radius and $1 \leq \gamma \leq 2$) and non cuspy
(i.e., with a flat density distribution near the center) can be built in this
way. The orbital structure of such models has been the subject of several
investigations over the past decade: see, e.g, \citet*{VKS02,MCW05,M06,AMNZ07}
for non cuspy models; or \citet*{MNZ09,ZM12} for cuspy ones; some authors
\citep{KV05,K08} even considered the effects of central black holes on the
orbital structure of their systems. All of these investigations found large
fractions of chaotic orbits in their models, and those fractions were
particularly high for the cuspy models and those with central black holes.

Nevertheless, \citet{HMSH01} starting from a spherical \citet{H90} model which
they adiabatically deformed to obtain a triaxial system, only found a negligibly
small fraction of chaotic orbits. It is worth noting, however, that \citet{KS03}
have already suggested that \citet{HMSH01} might have missed many chaotic orbits
due to the algorithm they employed to detect them. Besides, one should recall
that cold collapses necessarily result in models with strongly anisotropic
velocity distributions, because strongly radial orbits are produced by those
collapses, while the model of \citet{HMSH01} has a much more isotropic velocity
distribution (see their fig. 5, lower right). Now, chaotic orbits with low
angular momenta had already been found in models of our Galaxy by \citet{M74}
and, since then, further evidence that low angular momentum favors the onset of
chaos has been accumulating; for example, fig. 7 of \citet{MF96} shows that much
higher fractions of chaotic orbits are found among their "stationary start
space" orbits than among their "$x-z$ start space" ones. Therefore, it is also
possible that the difference in velocity anisotropy may help to explain the
different chaotic content of the collapse models and that of \citet{HMSH01}.

While the use of an $N$--body code might seem to guarantee the obtention of stable models,
that is not so, and the constancy of their macroscopic properties over intervals of the order
of a Hubble time should always be checked. Although \citet{HMSH01} investigated the stability of
their model, they did it over a time interval that was only a fraction of the Hubble time, so
that a check over a longer interval is warranted.

Thus, we decided to reinvestigate a model similar to the one of \citet{HMSH01}, checking its
long term stability and obtaining the fraction of chaotic orbits with different techniques
(section 2). Besides, we chose one of the models from the first paper in this series \citep{ZM12}
to "isotropize" its velocity distribution, rendering it much less radially oriented, and we also
adiabatically deformed the resulting system to obtain two additional ones; we subsequently obtained
the fraction of chaotic orbits in the three isotropized models (section 3). Section 4 of the
present paper presents our conclusions.

\section{Investigation of an adiabatically deformed model}
\subsection{The model}
\label{herntri}
The first part of our work investigates the stability of a model similar to that of \cite{HMSH01}
and its chaotic content. To build the model we began by setting up an isotropic Hernquist model
\citep{H90}  from its phase space distribution function:
\begin{eqnarray}
f({\bf x},{\bf v})=
\frac{M}{8\sqrt{2}\pi^3 a^3 w^3}\frac{1}{(1-s^2)^{5/2}} \nonumber \\
\times\left[ 3\arcsin s + s(1-s^2)^{1/2}(1-2s^2)(8s^4-8s^2-3)\right],
\end{eqnarray}
where $({\bf x},{\bf v})$ is a point of the phase space, $M$ is the total mass
of the model, $a$ stands for its length scale, $w=(GM/a)^{1/2}$, $G$ is the
gravitational constant, and $s=(-E)^{1/2}/w$, being $E=\frac{1}{2}v^2+\Phi(r)$
the energy per unit mass of a star in the potential $\Phi$ given by:
\begin{equation}
\Phi(r)=-\frac{GM}{r+a}.
\end{equation}
In these equations we have used the usual notations $r=|{\bf x}|$ and $v=|{\bf
v}|$. The density profile turns out to be
\begin{equation}
\rho(r)=\frac{Ma}{2\pi r}\frac{1}{(r+a)^3}.
\end{equation}
Since the density diverges as $r^{-1}$ at the origin, we say that the profile
has a cusp of (inverse) slope $\gamma=1$.

We set $G=M=a=1$ as \citet{HMSH01} did, yielding an overall crossing time of
$T_{\rm cr}=14$ time units (t.u. hereafter). We then generated $N=10^6$
particles according to the above distribution.

To obtain a triaxial system, we began by evolving the Hernquist model using the
self-consistent field (SCF) code of
\citet{HO92} with $n=6$ radial terms (the first one of which reproduces the
spherical Hernquist profile) and $l=10$ angular terms and, at the same time,
squeezing the system along the $z$ axis while maintaining its axial symmetry by
setting to zero all odd terms of the SCF expansion. The squeezing was done by
dragging the $z$ component of the velocities with the recipe of \citet{HMSH01}:
\begin{equation}
v_z'=v_z\frac{1-0.5\xi\Delta t}{1+0.5\xi\Delta t},
\end{equation}
where $\Delta t$ is the time step of the integration and $\xi$ is the
squeezing factor given by:
\begin{equation}
\xi=\xi_0\left[ 3\left( \frac{t}{t_{\rm gr}}\right)^2-
2\left( \frac{t}{t_{\rm gr}}\right)^3\right],
\end{equation}
where $t$ is the time elapsed since the beginning of the dragging, $t_{\rm gr}$ is
the interval during which the drag grows, and $\xi_0$ is an overall factor.
After the $z$-dragging reached its full strength, the system was evolved for a time
$t_{\rm drag}$ and, then, the drag was smoothly turned off by means of
\begin{equation}
\xi=\xi_0\left[ 1-3\left( \frac{t}{t_{\rm de}}\right)^2+
2\left( \frac{t}{t_{\rm de}}\right)^3\right],
\end{equation}
where $t_{\rm de}$ is the time of decay of the drag. Except for $\Delta t$, whose
value is not given by \citet{HMSH01} and that we took as $\Delta t=T_{\rm cr}/1600$, we
adopted the same values used by them, that is: $t_{\rm gr}=t_{\rm de}=10$ t.u.;
$t_{\rm drag}=30$ t.u., and $\xi_0=0.030 ($t.u.$)^{-1}$. 

Once the $z$ drag was over, the system was rescaled in positions and velocities in
order to recover the original value of the energy. We then applied the same procedure
just described to squeeze the system along the $y$ axis, and the system was
rescaled once again to recover the initial energy value. For the $y$ dragging the
requirement of axisymmetry was lifted and $\xi_0$ was taken as $0.025 ($t.u.$)^{-1}$,
again, the same value used by \citet{HMSH01}. Finally, following their procedure,
we let the system evolve until $t=180$ t.u. (corresponding to about $5.7 T_{\rm cr}$)
to reach an equilibrium state; for this evolution, as well as the longer one described
below, the timestep of integration was taken as $T_{\rm cr}/800$.

\begin{figure*}
\resizebox{\hsize}{!}{\includegraphics{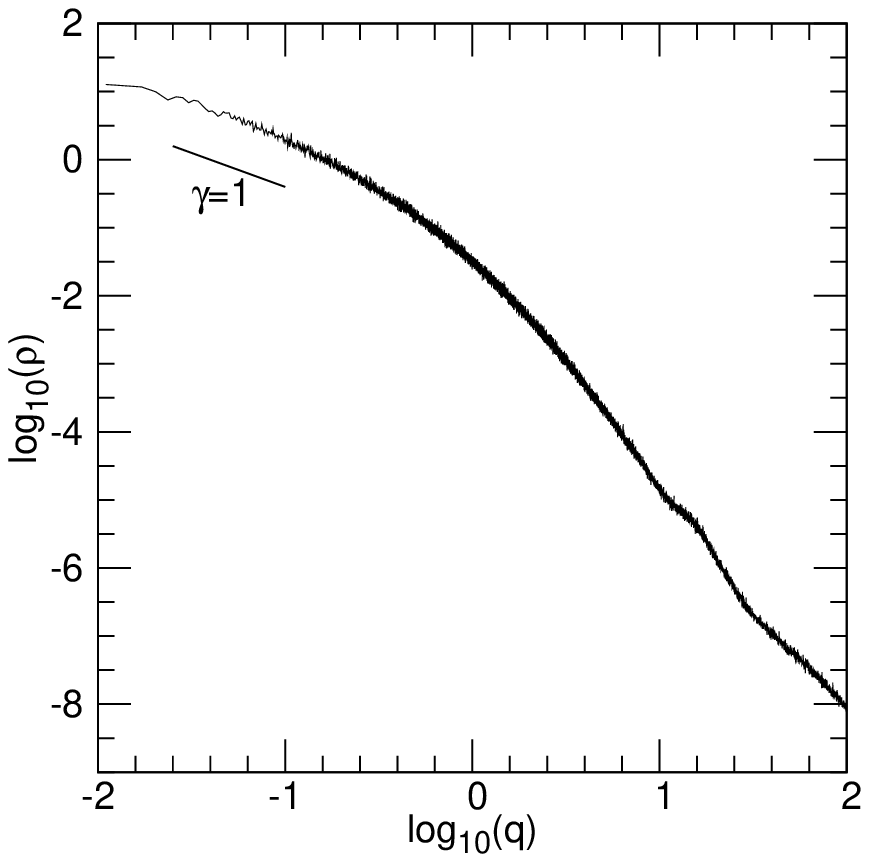}\hspace{1cm}
                      \includegraphics{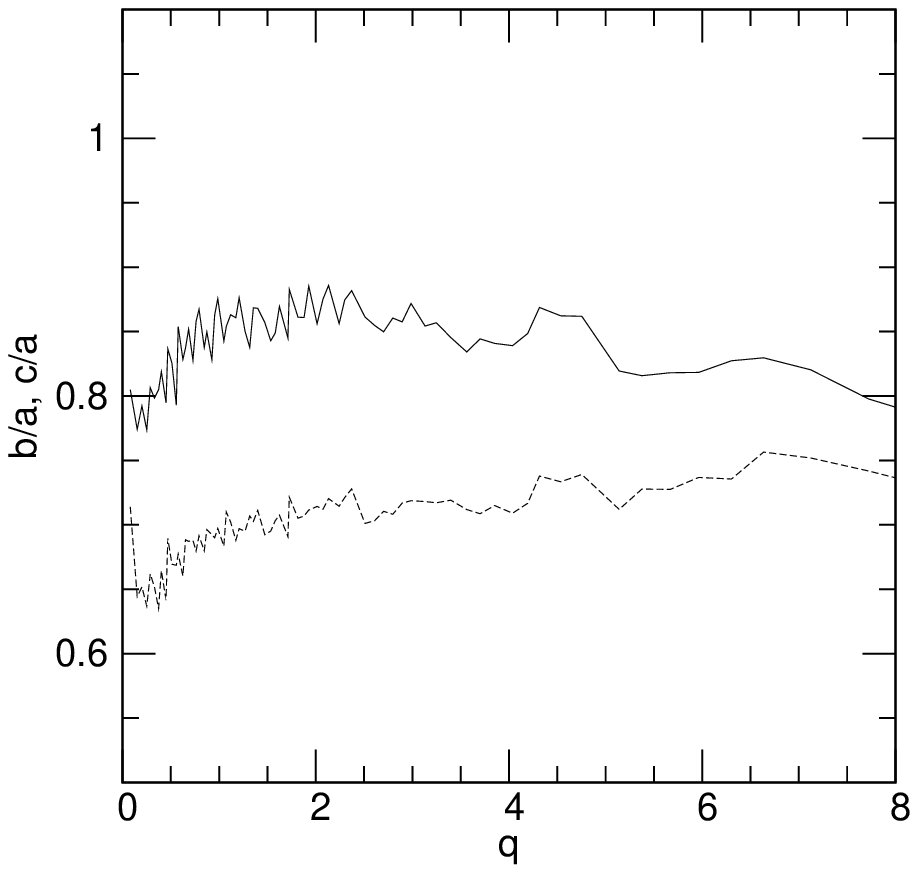}}
\resizebox{\hsize}{!}{\includegraphics{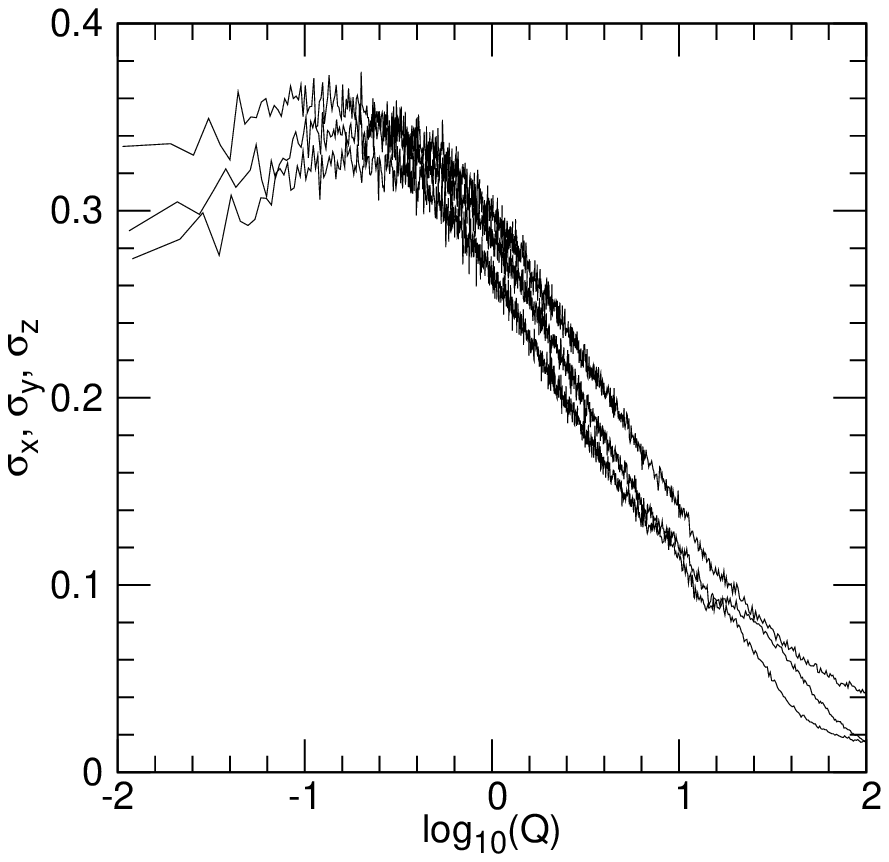}\hspace{1cm}
                      \includegraphics{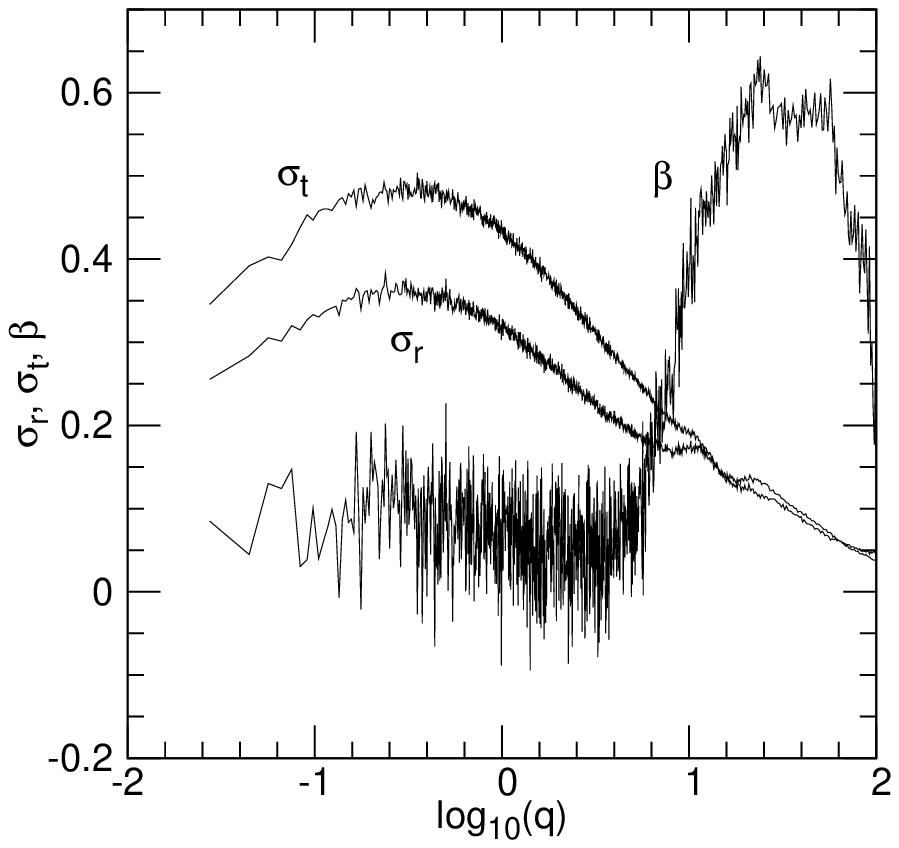}}
\caption{Structural and kinematical properties of our model; $q$ is the ellipsoidal radius and $Q$ the
projected ellipsoidal radius. Upper left: density profile, a short
segment with slope $-1$ was added for reference. Upper right: axial ratios, the upper curve corresponds
to $b/a$ and the lower one to $c/a$. Lower left: projected velocity dispersions along the main axes,
upper, intermediate and lower curves correspond, respectively, to $\sigma_{x}, \sigma_{y}$ and $\sigma_{z}$.
Lower right: tangential and radial velocity dispersions and velocity anisotropy parameter $\beta$.}
\label{como5deHBea}
\end{figure*}

The properties of the final model are presented in Fig. \ref{como5deHBea} which
was prepared in the same way as fig. 5 of \citet{HMSH01} to underline the strong
similarity of their model and ours. Just as they did, we use the ellipsoidal radius:
\begin{equation}
q=\left( \frac{x^2}{a^2}+\frac{y^2}{b^2}+\frac{z^2}{c^2} \right)^{1/2},
\end{equation}
with $a=1$,
and its two dimensional projection, $Q$. The semiaxes, $a, b$ and $c$ were computed following
the recipe of \citet{DC91}, as adapted by \citet{HMSH01}. $\langle v_{\rm t}^2\rangle$
and $\langle v_{\rm r}^2\rangle$ are the mean square tangential and radial velocities,
respectively, $\sigma_{\rm t}$ and $\sigma_{\rm r}$ are the corresponding velocity dispersions,
and $\beta$ is the anisotropy parameter, defined as usual as
$\beta=1-\langle v_{\rm t}^2\rangle / (2\langle v_{\rm r}^2\rangle)$. The density and the velocity
dispersions were computed using constant axial ratios evaluated at the half-mass radius.
A comparison of
our Fig. \ref{como5deHBea} with fig. 5 of \citet{HMSH01} clearly shows that
both systems are very similar indeed.

\begin{figure}
\includegraphics[width=\hsize]{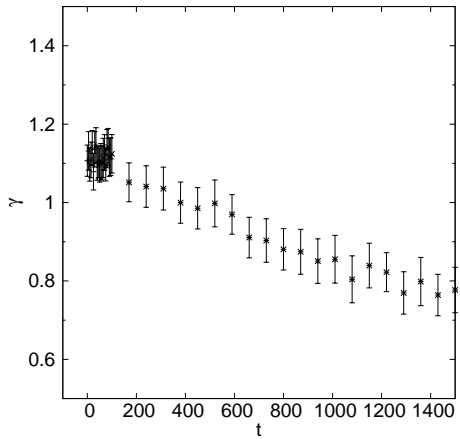}
\caption{Time evolution of the slope of the central density $\gamma$, when the
system is let to relax $100 T_{\rm cr}$.}
\label{gammatl}
\end{figure}

The time evolutions of the central slope, $\gamma$, and of the $b$ and $c$ semiaxes
of our model (not shown here) are also very similar to those shown by \citet{HMSH01}
in their figs. 2 and 3, but we should recall that the time interval that they let
their model relax was only 80 t.u., i.e., $5.7 T_{\rm cr}$. Therefore, we
let our system relax for an additional interval of $t=1,400$ t.u.,i.e., about
$100 T_{\rm cr}$, in order to verify whether it had truly reached
equilibrium. Fig. \ref{gammatl} shows the resulting evolution of the central slope
computed from the innermost 10,000 particles divided in 100 particles bins, and it
is clear from the figure not only that the system had not yet achieved equilibrium
at $t=180$, but also that the central cusp flattens strongly, falling to
$\gamma < 0.8$ after a $100 T_{\rm cr}$ evolution.

\begin{figure}
\includegraphics[width=\hsize]{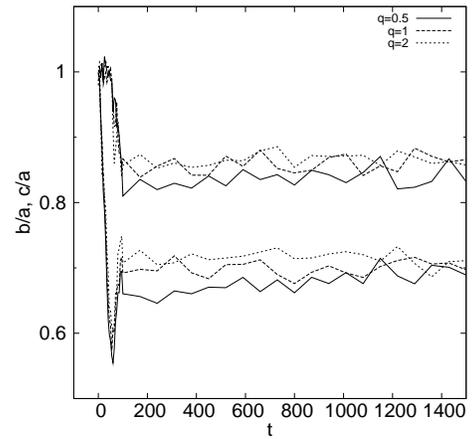}
\caption{Time evolution of the axial ratios, for three different
values of $q$, when the system is let to relax $100 T_{\rm cr}$.}
\label{semistl}
\end{figure}

\begin{figure}
\includegraphics[width=\hsize]{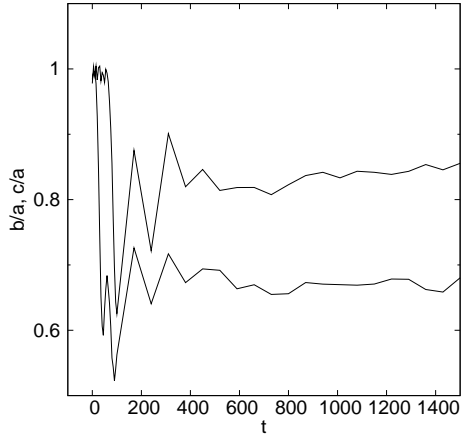}
\caption{Time evolution of the axial ratios of the 80 per cent most bounded
particles, when the system is let to relax $100 T_{\rm cr}$.}
\label{semis80tl}
\end{figure}

Alternatively, when we computed the evolution of the semiaxes corresponding
to $q=0.5, 1$ and 2, we found that they maintained reasonably well the values
reached at $t=180$ t.u. (Fig. \ref{semistl}). Nevertheless, the evolution
of the semiaxes of a further ellipsoidal slice of the system, the one which
encompasses 80 per cent of the most tightly bound particles, corresponding to
about $q=8$ at the start of the simulation, (Fig. \ref{semis80tl}) shows
that these semiaxes are strongly out of equilibrium until at least $t=500$ t.u.
Thus, although in order to compare our results with those of \citet{HMSH01},
we will investigate in what follows the chaoticity of the model obtained from
the $t=180$ t.u. snapshot, it should be borne in mind that the system has
not reached equilibrium and is still evolving.

\subsection{Chaos detection with frequency differences}

In order to establish the fraction of chaotic orbits present in the system,
we began by repeating the procedure of \citet{HMSH01}. We selected at random
5,000 particles from the $t=180$ t.u. snapshot, and used them as initial conditions
to integrate orbits. The integrations were performed in the potential generated
by the SCF code using the complete collection of particles to which, following
 \citet{HMSH01} and in order to reduce the noise in the computation of the
potential, we added seven additional sets of particles in each of which
the particles were replicated in a different octant in coordinates as well
as in velocities. 

Then we followed each orbit with a Runge-Kutta-Fehlberg integrator of order 7/8,
over a time interval of 900 orbital periods, and we used the coordinates and velocities
of the first and the last 300 period intervals to obtain the dominant frequencies of the
orbit in those intervals, $f_1$ and $f_2$, respectively. Those frequencies were obtained
with the modified Fourier transform algorithm, first developed by \citet{L88} and later
on improved by \citet{SN97}, using 8,192 points for each 300 period interval. The complex
variables we used had the coordinates as their real part and the corresponding velocity
component as the imaginary part. Of the three resulting frequencies (one for each
coordinate) we chose for the analysis the one that corresponded to the largest amplitude
in the first 300 period interval. If the orbit were regular, its fundamental frequency
should not change, i.e., $f_1=f_2$; on the contrary, a chaotic orbit should show
$f_1\ne f_2$. In order to perform the comparison of $f_1$ with $f_2$ and determine
if they are different (in a numerical sense), we followed the  recipe of
\citet{HMSH01} (see also \citet{VM98}), that is, we checked whether
\begin{equation}
\Delta f=\frac{|f_1-f_2|}{f_0 T} > 0.05\left( \frac{300 {\rm periods}}
{900 {\rm periods}}\right)^{1/2}\simeq 0.0289,
\label{deltaf}
\end{equation}
where $f_0$ is a frequency of reference defined in \citet{HMSH01} as "the
frequency of a  tube orbit about the long axis", $T$ is defined  as "the time
interval", and the constant inside the square root is our equivalent of the
constant used by \citet{HMSH01}. As $\Delta f$ is compared against a constant to
determine chaoticity, we felt  that the reference frequency $f_0$ should  not be
that of just any long axis tube. Thus, we first chose to compute $f_0$ from the
long axis tube with the same energy of the orbit being classified. But this led
to a rather time-consuming and complicated algorithm, so we shifted to the
original reference frequency defined in \citet{VM98}, i.e., that of the long
axis orbit with the same energy of the orbit being classified. As these orbits
pass exactly through the origin, and the potential diverges there, we sligthly
perturbed the initial conditions in order to avoid that point. With respect to
$T$, we assumed that it corresponds to the time span from the first sample of
the orbit used to compute $f_1$ to the last sample of the orbit used to compute
$f_2$.

\begin{figure}
\includegraphics[width=\hsize]{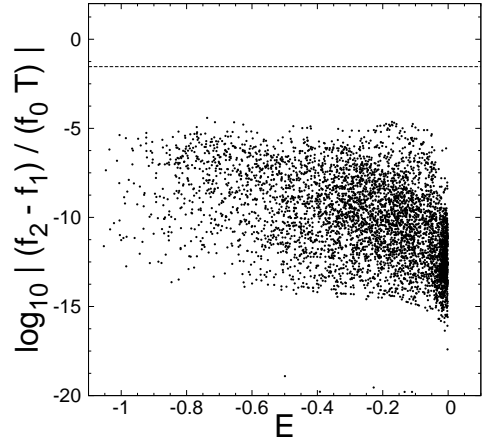}
\caption{Frequency differences $\Delta f$, Eq. (\ref{deltaf}), of the 5,000
selected orbits, as a function of orbital energy.}
\label{dft}
\end{figure}

We found that, according to this criterium, none of the orbits was chaotic. Moreover,
the values of $\Delta f$ lay in all cases well below the threshold value (Fig.
\ref{dft}). Let us note that a) there is no clue as to where the limit
between regular and chaotic orbits may be; b) if the (arbitrary?) factor 0.05 in
Inequality  (\ref{deltaf}) were changed to another, much lower, value chaos
would have been found; and c) a dimensional analysis of Inequality (\ref{deltaf})
reveals that the first member has units of time$^{-1}$, whereas  the second
member is a dimensionless constant. Thus, Inequality (\ref{deltaf}) seems to be
problematic.

We therefore explored the possibility of detecting chaos with other indicators based
on frequency differences. We used $\Delta f$ without the $T$ factor:
\begin{equation}
\Delta f'=\frac{|f_1-f_2|}{f_0},
\label{deltaf1}
\end{equation}
the relative difference, but referred to the original frequency:
\begin{equation}
\Delta f''=\frac{|f_1-f_2|}{f_1},
\label{deltaf2}
\end{equation}
and the absolute difference between frequencies,
\begin{equation}
\Delta f'''=|f_1-f_2|,
\label{deltaf3}
\end{equation}
without imposing any a priori threshold. The plots of $\Delta f'$, $\Delta f''$ and
$\Delta f'''$ versus the orbital energy turned out to be very similar to that of
Fig. \ref{dft}, giving no clue as to where to set a limiting value separating regular from
chaotic orbits. Therefore, the question remains: Where is to be put the limit between regular
and chaotic orbits, if there is such a limit at all? We cannot answer for sure until we
manage to get some additional evidence from another point of view.

\subsection{Chaos detection with Lyapunov exponents and SALI}

\begin{figure}
\includegraphics[width=\hsize]{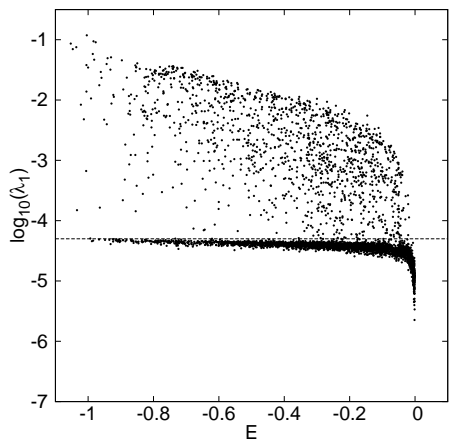}
\caption{Largest FT-LCNs of the chosen 5,000 orbits as a function of orbital energy. The
threshold value $\Lambda=5\times 10^{-5}$ is also shown.}
\label{lyapu}
\end{figure}

We decided to analyse the chaotic content of the model using
Lyapunov exponents which constitute, numerical issues aside, the very
definition of chaos. As in our previous papers, we used the LIAMAG routine \citep{UP88}
to compute Finite Time Lyapunov Characteristic Numbers (FT-LCNs). We took the same random
sample of 5,000 initial conditions from the $t=180$ t.u. system we had used for the
frequency analysis and we integrated the orbits in the potential
generated by the SCF over $t=$280,000 t.u. (i.e., 20,000 $T_{\rm cr}$); the
renormalization interval was set as 28 t.u.
Fig. \ref{lyapu} shows the largest FT-LCN, $\lambda_1$, as a function of the energy
of the orbits and, while Fig. \ref{dft} gave no hint about where to
place a limit separating regular from chaotic orbits, here there is a clearly
defined limit at $\Lambda=5\times 10^{-5}$. The same limiting value can be found from a
similar plot for the second positive FT-LCN, and one can use it to refine the
classification by separating chaotic orbits into totally chaotic (having the two
largest FT-LCNs greater than $\Lambda$) and partially chaotic (having only the
largest FT-LCN greater than $\Lambda$). The resulting classification was 73.1 per
cent of regular orbits and 26.9 per cent of chaotic orbits, of which 21.1 per
cent were totally chaotic and the remaining 5.7 per cent partially chaotic.
This is in sharp contrast with the less than 1 per cent chaotic content found by
\citet{HMSH01}, but not surprising considering our results with the frequency
difference technique that they had used.

\begin{figure}
\includegraphics[width=\hsize]{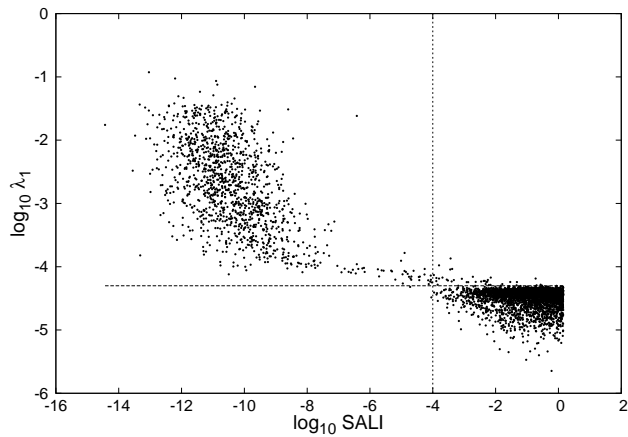}
\caption{Relationship between the largest positive FT-LCN and the SALI for our
sample set of orbits. The straight lines indicate the chosen thresholds of both
indicators. Only a handful of orbits are differently classified by each chaos
indicator.}
\label{lyasali}
\end{figure}

We checked the abovementioned results by computing another chaos indicator,
namely the SALI \citep{S01,SABV04}. Fig. \ref{lyasali} shows the relationship
between the largest FT-LCN and the SALI. We can see the typical bridge of this kind
of plots, joining regular and chaotic orbits \citep[cf.][their fig. 3]{VKS02}.
Taking SALI $=10^{-4}$ as a typical limiting value between regularity and chaocity, we
obtained 73.8 per cent of regular orbits and 26.2 per cent of chaotic ones. Of
course, the SALI does not allow further classification into fully and partially
chaotic orbits but it is clear that the FT-LCN and the SALI give essentially the
same classification into regular and chaotic orbits with only a handful of
exceptions.

\begin{figure}
\includegraphics[width=\hsize]{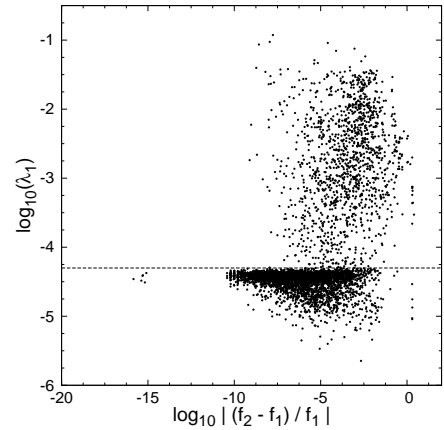}
\caption{Frequency differences $\Delta f''$, Eq. (\ref{deltaf2}),
versus the maximal FT-LCNs for the sampled 5000 orbits.}
\label{lyadif}
\end{figure}

Compared to the excellent agreement between the FT-LCN and the SALI, the correlation
between any of the frequency difference indicators and the FT-LCN is very poor, as
shown by Fig. \ref{lyadif}, which presents $\Delta f''$ versus the largest FT-LCN,
$\lambda_1$, for our sample of orbits (similar plots were obtained with $\Delta f$,
$\Delta f'$ and $\Delta f'''$, and are not shown). The only good agreement that can be
found is for low values of the frequency differences, that correspond also to low values
of the FT-LCNs.  For larger values of the frequency differences, instead, we found
essentially a vertical cloud of points where, for the same value of frequency diference,
coexist FT-LCN values that correspond to regular and to chaotic orbits, and no correlation
between frequency differences and FT-LCNs is apparent within the cloud either. Again,
no clear separation between low and high frequency difference values is apparent and
any vertical line that one traces will render an arbitrary limit only. It is worth
recalling that the poorer perfomance of frequency differences as compared to other chaos
detectors had already been noted by \citet{KV05} and by \citet{MDCG13}. This is probably
one reason why the results of \citet{HMSH01} differed so much from those of others which
used FT-LCNs or SALIs as chaos detection techniques.

\begin{figure}
\includegraphics[width=\hsize]{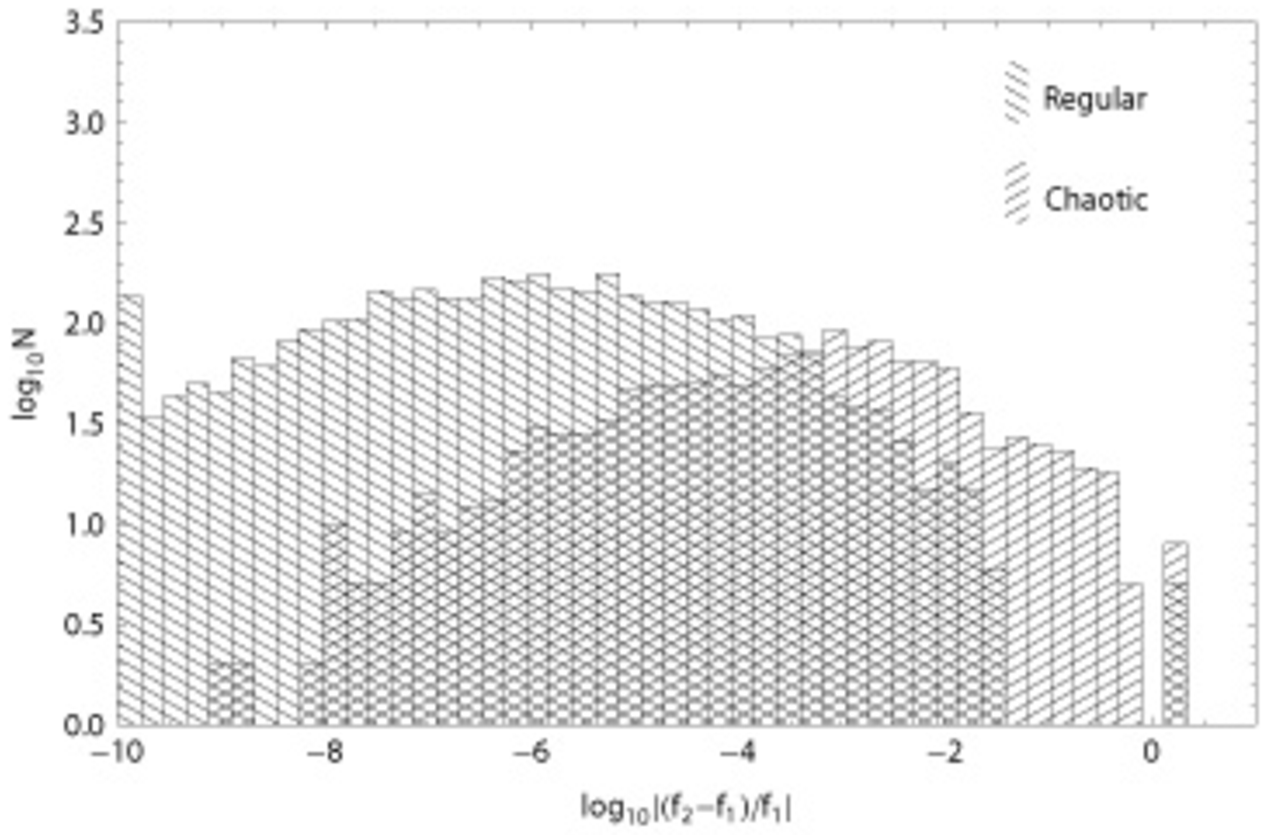}
\includegraphics[width=\hsize]{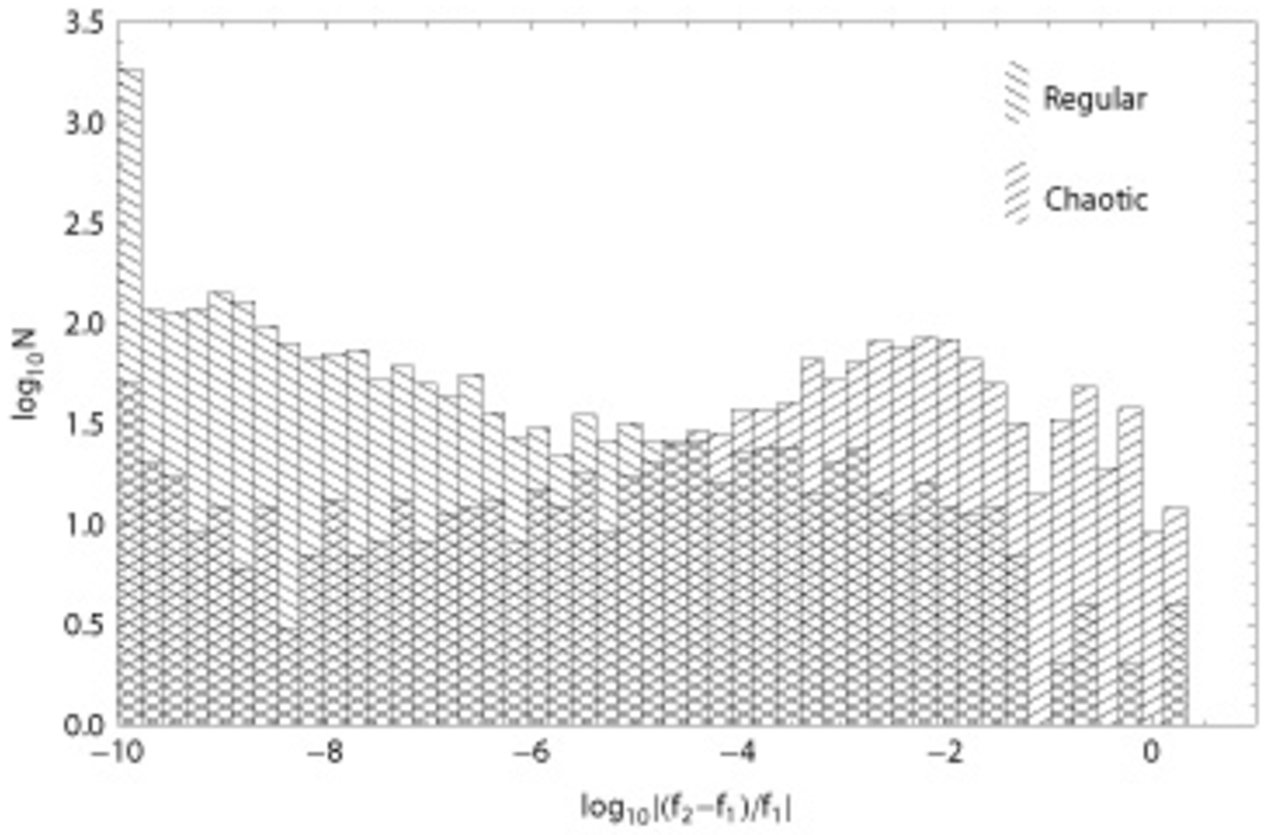}
\caption{Histograms of frequency differences, separating the regular
and chaotic orbits according to the FT-LCNs, using 300 (above) and 4,800 periods (below).
In both diagrams the first left column corresponds to all the frequency differences whose
logarithm is smaller than $-9.75$.}
\label{histo}
\end{figure}

To be fair, it should be recognized that the computing times of the
frequency differences are much shorter than those of the FT-LCNs or the SALIs. In our case,
for example, the computation time for the frequency differences was about 4.5 times shorter
than that for the FT-LCNs. Nevertheless, although increasing the number of periods and the
time interval to 4,800 periods (i.e., 16 times 300) does improve the results of the method of
frequency differences, the improvement is still not enough to provide a good separation of
regular from chaotic orbits. Fig. \ref{histo} presents the histograms of the frequency
differences for 300 (above) and 4,800 periods (below), separately for the orbits shown to
be regular and chaotic by the FT-LCNs. Even though the computation time for the frequency
differences with 4,800 periods was about 3.5 times that for the FT-LCNs, they still do not
allow a clear separation of regular from chaotic orbits. The large number
of regular orbits that have high frequency differences is probably due to the presence of
nearby lines that are difficult to separate with the Fourier technique, but the increased
number of chaotic orbits with very low frequency differences is more puzzling. Perhaps
the very large intervals used to compute the frequencies (4,800 periods) contribute to
average the frequency changes that can be expected from those orbits. More details on the
use of the frequency differences for chaos detection can be found on the recent paper
by \cite{V13}, but it should be stressed that our main interest was just to show that their
use may have caused the extremely low fraction of chaotic orbits found by \citet{HMSH01},
rather than providing a full analysis of the advantages and disadvantage of the different
chaos indicators. 

\citet{HMSH01} follow \citet{VM98} in arguing that orbits with low frequency differences,
although chaotic, have diffusion times too slow to substantially alter the shape of the
system over time intervals of the order of a Hubble time but, in fact, the system can
be highly stable even when containing strongly chaotic orbits, as was shown by \citet{ZM12}
and as we will show again later on in the present paper. Moreover, quite a few orbits
have FT-LCNs in excess of $0.01 ($t.u.$)^{-1}$, that is, Lyapunov times shorter than
100 t.u. and, since $T_{\rm cr}=14$ t.u., they can hardly be called weakly chaotic.

From the point of view of galactic dynamics what really matters to distinguish orbits in
a model is whether they have the same distribution or not and, to investigate that issue,
Table \ref{semiorb} gives the root mean squared values of each Cartesian coordinate,
computed from eleven orbital points of every orbit in each set, and the corresponding
ratios together with their errors. As we have shown before \citep{MCW05,AMNZ07},
regular, partially and fully chaotic orbits in non cuspy systems have different degrees of
flattening. For the cuspy models of \citet{ZM12} those differences were less obvious, as
they are for the present model as shown in columns five and six of the Table. The likely
cause is the presence of the cusp itself, that should induce more chaotic behaviour
in the orbits that come nearer to it. That is confirmed by columns two, three and
four of the same Table, where we see that regular orbits have a much more extended
distribution than partially chaotic orbits and that these have a more extended
distribution than fully chotic orbits. That is, from the point of view of galactic
dynamics the separation of the orbits we have made here is highly relevant.

\begin{figure}
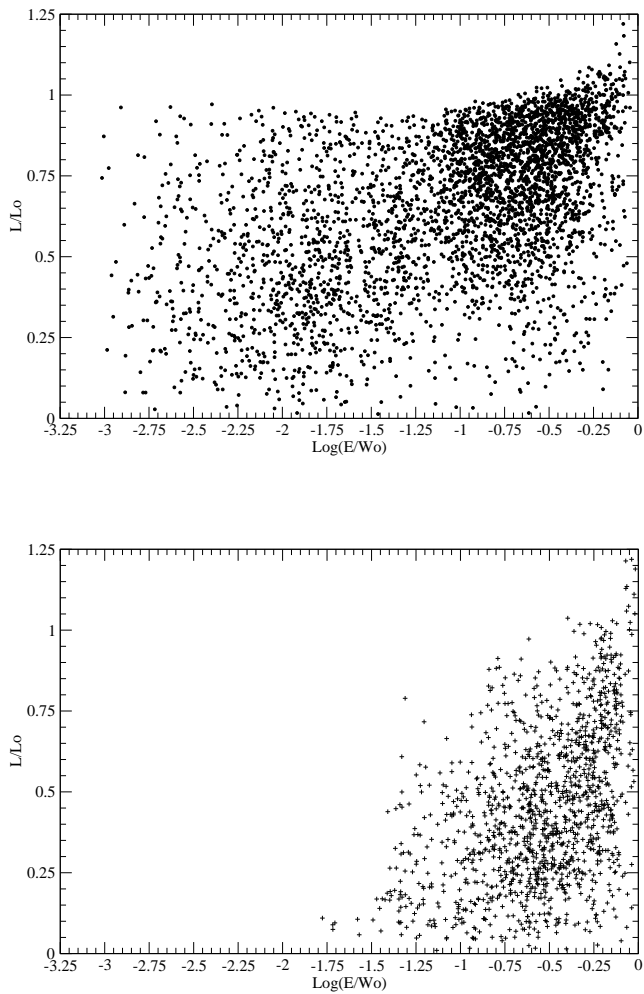

\vskip 5mm
\includegraphics[width=\hsize]{HBetalreg.eps}
\vskip 11mm
\includegraphics[width=\hsize]{HBetalchao.eps}
\caption{Reduced energy versus reduced initial angular momentum for the regular (above) and
chaotic (below) orbits of the sample of 5,000 orbits.}
\label{HBea}
\end{figure}

Fig. \ref{HBea} presents, for the regular (above) and chaotic (below)
orbits, the plots of their reduced energy versus their reduced initial angular
momentum, the normalizing factors being the potential energy at the center of
the system and the maximum angular momentum, respectively; the latter was
computed assuming circular orbits in a Hernquist potential. Both partially and
fully chaotic orbits were bunched together, as chaotic, for clarity. The left
portion of the energies (i.e., energies close to zero) has exclusively regular
orbits, in agreement with the results of Table \ref{semiorb}; on the right part,
its is clear that the chaotic orbits tend to have lower angular momenta than the
regular orbits of the same energy, confirming that low angular momenta favor the
onset of chaos.

\begin{table*}
 \caption{Geometrical distribution of regular, partially chaotic and fully
 chaotic orbits of our sample.}
 \label{semiorb}
 \begin{tabular}{lccccc}
  \hline
  Type & $a=\langle x^2\rangle^{1/2}$ & $b=\langle y^2\rangle^{1/2}$ 
       & $c=\langle z^2\rangle^{1/2}$ & $b/a$ & $c/a$ \\
  \hline
  Regular & $47.96\pm0.94$ & $44.92\pm0.85$ & $45.09\pm0.86$ & $0.936\pm0.025$ & $0.940\pm0.026$\\
  Part. chaotic & $4.40\pm0.16$ & $3.58\pm0.17$ & $3.72\pm0.19$ & $0.814\pm0.048$ & $0.847\pm0.053$ \\
  Fully chaotic & $2.24\pm0.04$ & $1.92\pm0.07$ & $2.02\pm0.07$ & $0.858\pm0.034$ & $0.904\pm0.035$ \\
  \hline
 \end{tabular}
\end{table*}

\section{Isotropized models}
\label{velotropo}
We have now established that there is a significant fraction of chaotic orbits
in a model similar to the one of \citet{HMSH01} and that the reason why they did
not find them is, in all likelihood, their use of frequency differences as chaos
detector. Nevertheless, the chaotic fraction found in our model is
much lower than the fractions typically found in models built from cold
collapses and, as indicated in the Introduction, it is possible that the strongly
radial orbits resulting from those collapses are the key to explain that
difference. Besides, as we have shown, the model we built following the recipe of
\citet{HMSH01} had not reached equilibrium, so that the question remains whether it
is possible to build cuspy triaxial models whose velocity distributions are not
radially biased, that include significant fractions of chaotic orbits and are highly
stable. Therefore, we decided to try to render more isotropic the velocity distribution
of a model obtained from a cold collapse and to see whether its chaotic fraction
decreases as a result and the final model is stable. Since the models from the
first paper of this series \citep{ZM12} are, to our knowledge, the ones with the
highest fractions of chaotic orbits reported, with less than 25 per cent of
regular orbits, they seem to offer a good starting point.

\begin{figure*}
\vspace{20pt}
\centering
\begin{minipage}{160mm}
\includegraphics[width=7.5truecm]{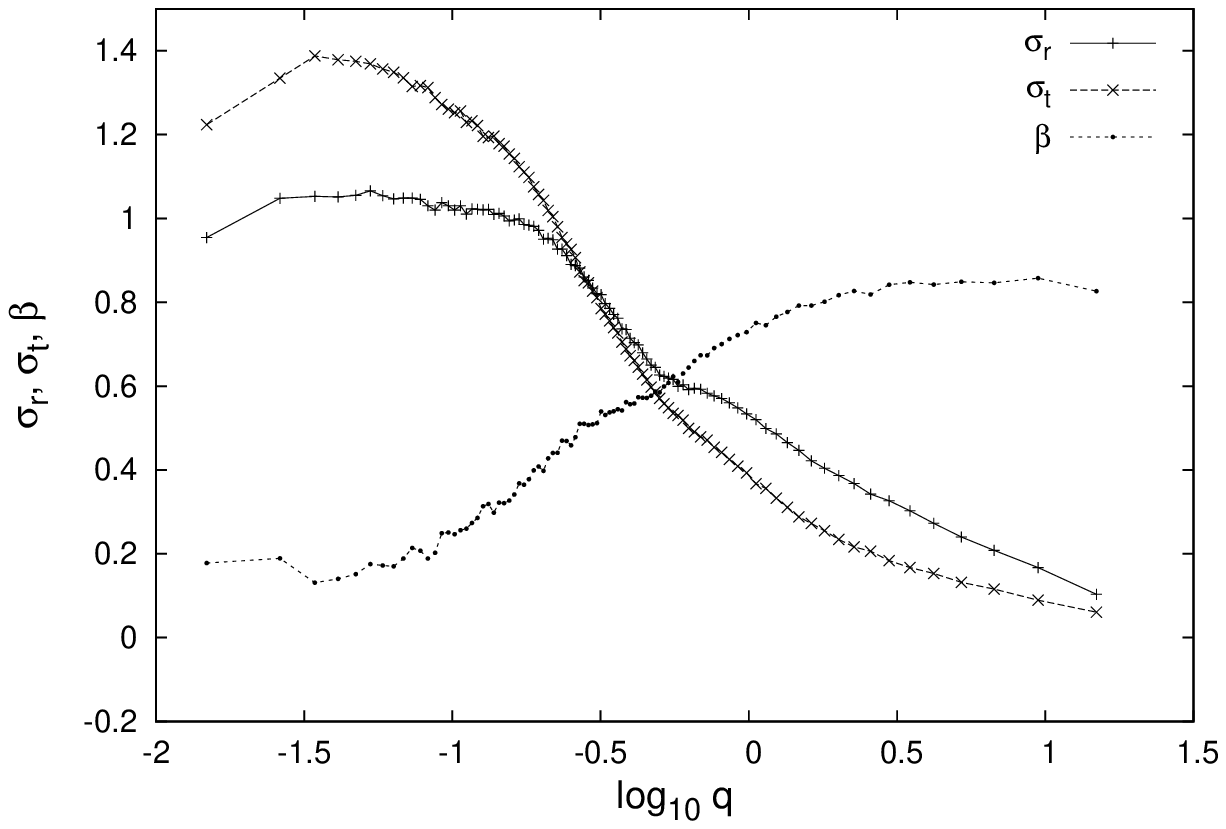}~\hfill\includegraphics[width=7.5truecm]{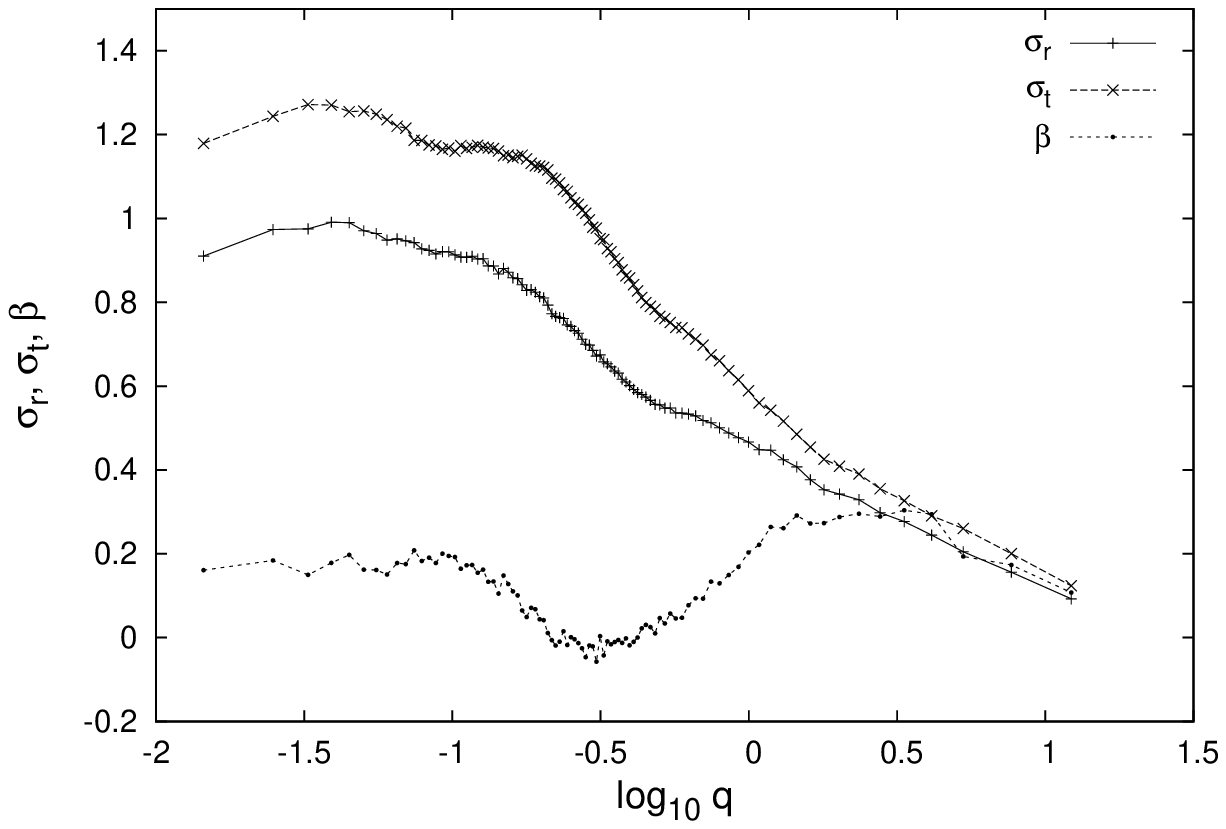}
\vskip 10.5mm
\includegraphics[width=7.5truecm]{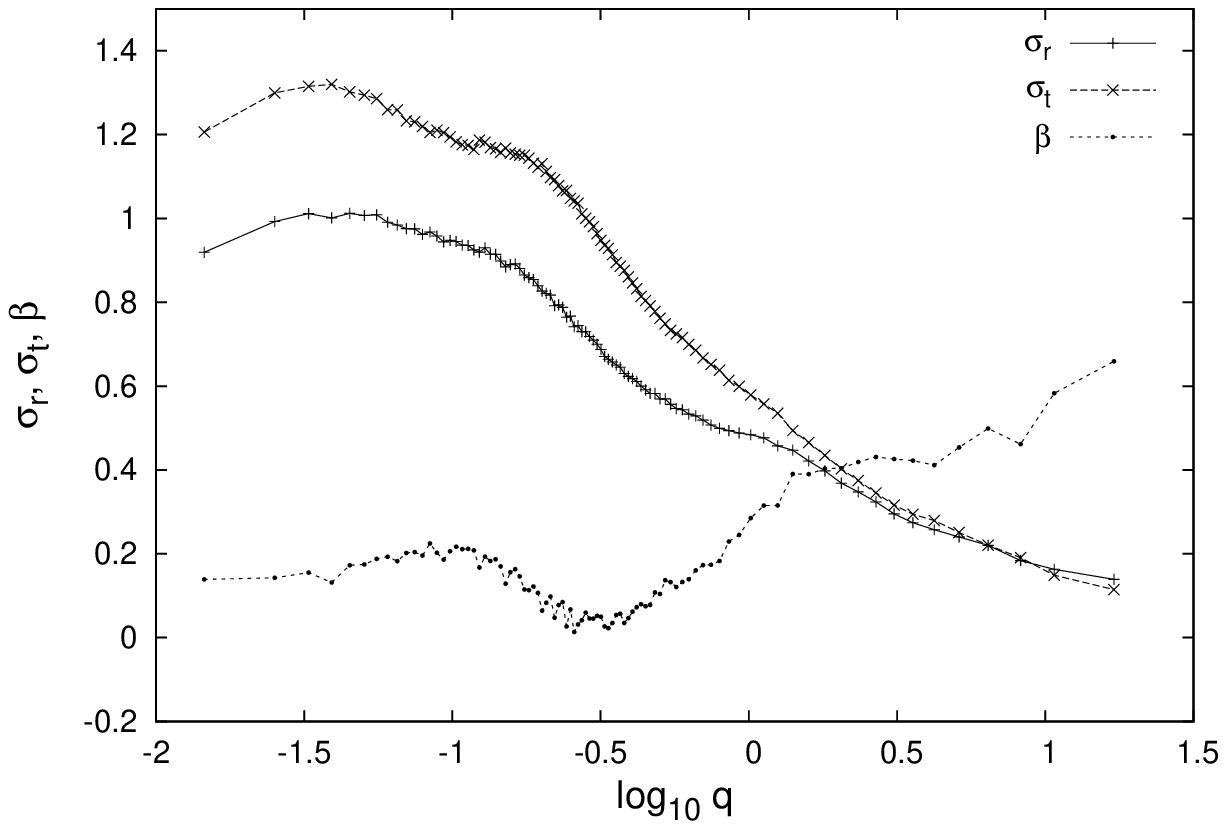}~\hfill\includegraphics[width=7.5truecm]{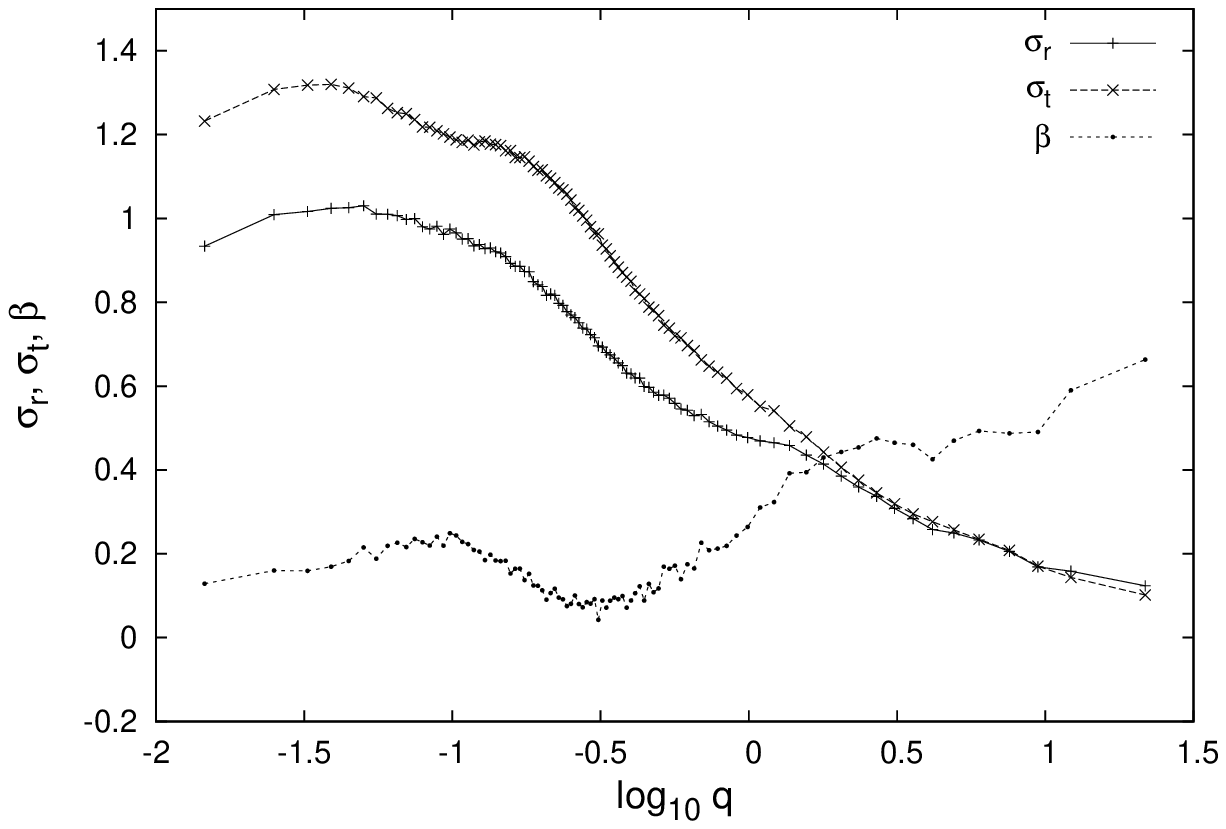}
\caption{Radial and tangential velocity dispersions, and velocity anisotropy
parameter $\beta$, for models: E4c of Paper I (upper left); E4ci (upper right);
E4cf6 (lower left); E4cf8 (lower right).}
\label{betae4c}
\end{minipage}
\end{figure*}

We found that isotropization tends to flatten somewhat the central cusp (see below),
so that we chose the model with the steepest central cusp ($\gamma=1.073\pm 0.022$)
dubbed E4c which, with only 12.67 per cent regular orbits, is also one of those with
highest chaotic content. This model has semiaxes ratios $b/a=0.765$ and $c/a=0.575$,
($a>b>c$ are the square roots of the mean square values of the $x, y, z$ coordinates,
respectively, of the 80 per cent most tightly bound particles) and triaxiality
$T=(a^2-b^2)/(a^2-c^2)=0.62$. The overall crossing time for this model is
$T_{\rm cr}=0.521$ t.u. Fig. \ref{betae4c} (upper left) shows the velocity dispersions in
the radial and tangential directions, as well as the velocity anisotropy parameter
$\beta$ for this model. A comparison with Fig. \ref{como5deHBea} (lower right) should
take into account the different scales, so that longitudes and velocities from that
figure should be multiplied by 0.097 and 2.6, respectively, to switch them to the
scales of Fig. \ref{betae4c}. With that caveat, it is obvious from the figures
that the $\beta$ values from model E4c are substantially higher than those of the model
of Section 2 for all ellipsoidal radii. 

\subsection{Isotropization technique}
\begin{figure}
\includegraphics[width=\hsize]{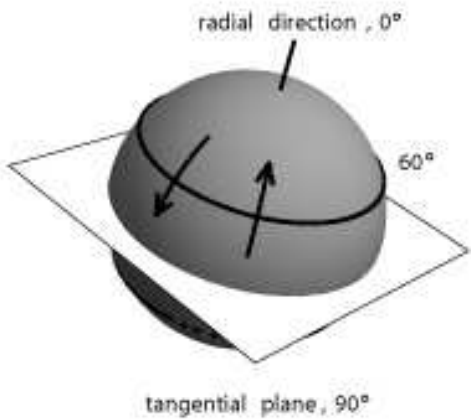}
\caption{Rotation of the velocity vectors. The spherical surface represents the
locus of the unit velocity vectors. The velocities which lie on the radial cap
rotate towards the tangential plane, and vice versa.}
\label{esfera}
\end{figure}

We began selecting at random 25 per cent of the particles from model E4c and, for each
particle, we computed the angle $\theta$ between the velocity vector and the radial
direction. If $0^\circ \le \theta\le 60^\circ$, then the vector was rotated on
the plane defined by the radius and the velocity towards the
tangential direction, until it lay in the region $60^\circ \le \theta\le
90^\circ$ (Fig. \ref{esfera}). The actual angle of rotation was computed by
linearly mapping the interval $[0^\circ,60^\circ]$ into the interval
$[60^\circ,90^\circ]$. The same was done if the velocity vector originally lay
in the interval $[120^\circ,180^\circ]$, in which case it was rotated into the
region $[90^\circ,120^\circ]$. Alternatively, when the velocity vector made
an angle between $60^\circ$ and $90^\circ$ with the radial direction, it
was rotated towards this last direction by mapping linearly from
$[60^\circ,90^\circ]$ to $[0^\circ,60^\circ]$. Finally, those velocity vectors
lying on $[90^\circ,120^\circ]$ were similarly rotated into the region
$[120^\circ,180^\circ]$. Since the $[0^\circ,60^\circ]$ cap around the radial
direction has the same area as the $[60^\circ,90^\circ]$ zone, then
approximately the same number of velocity vectors were rotated towards the
radial direction than towards the tangential direction, thus avoiding any
hollowing out of the velocity ellipsoid. But, since the models are originally
radially anisotropic, those velocities near the radial direction have bigger
moduli than those near the tangential plane and, thus, our procedure extracts
"velocity power" from the radial direction and puts it nearer the tangential
plane, i.e., the velocity ellipsoids are left more isotropic than before.

Once the above procedure was finished, we let the system relax for 10 t.u., which for
this model corresponds to about  $20 T_{\rm cr}$, to allow it
to reach a new equilibrium. For this evolution we used the code of \citet{HO92} with
$n=6$ radial terms and $l=4$ angular terms \citep{ZM12}. After the relaxation,
we took another random 25 per cent of particles (from those that had not been
chosen before) and repeated the velocity rotation algorithm, followed
by another 10 t.u. relaxation period. The rotation plus relaxation process was repeated
two more times, thus rotating the velocities of all the particles of the system
in the end. The resulting system was allowed to relax an additional $100 T_{\rm cr}$ to
obtain the final model, dubbed E4ci. It has a fairly isotropic velocity distribution
with $\beta\simeq 0$, except at large radii where $\beta > 0.2$ (see Fig. \ref{betae4c},
upper right). Nevertheless, the central cusp flattened somewhat during the isotropization
process, and the final model has $\gamma=0.899 \pm 0.019$, still a reasonably steep cusp.
We checked the stability of the system by letting it evolve another $400 T_{\rm cr}$
and verifying that its global parameters (central density and moments of inertia of
the 80 per cent most tightly bound particles) were conserved, just as we had done
for the models of \citet{ZM12}. The results were even better than those found there, with
all the changes smaller than 1 per cent in 100 t.u., i.e., about a Hubble time.
In addition to finding those changes from the self--consistent evolution
of their models, \citet{ZM12} performed two additional tests for one of them: 1) They
fixed the potential before letting the model to evolve (i.e., they eliminated self--consistency);
2) They took one tenth of the particles, and increased ten times their masses, and they let
the model evolve self-consistently. The results were that with the fixed potential the evolution
was much smaller, and with the reduced number of particles much larger, than for the original
model, so that they concluded that most of the evolution was simply due to numerical relaxation
in the Hernquist and Ostriker code. Similar tests were done by \citet{AMNZ07} and \citet{MNZ09}
using the Aguilar code and they obtained similar results and reached the same conclusion. Thus,
the small percentual changes found here are also most likely
due to relaxation effects of the $N$--body code and the stability of the model is
even better than those percentages indicate. Here we also checked the change
in the slope of the central cusp and it turned out to be negligibly
small, a mere $-0.03 \pm 0.99$ per cent in a Hubble time.

\begin{figure}
\vskip 11mm
\includegraphics[width=\hsize]{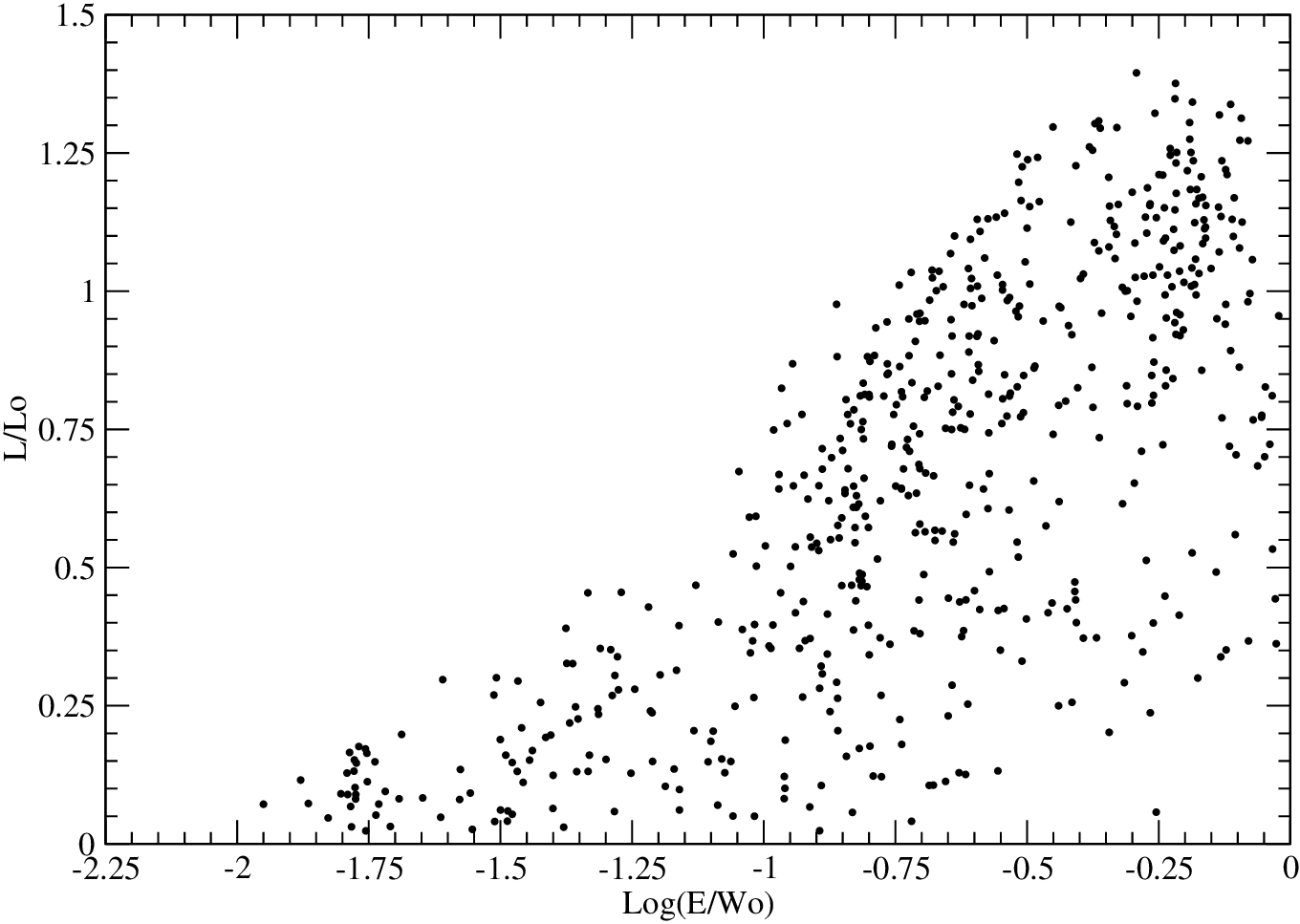}
\vskip 11mm
\includegraphics[width=\hsize]{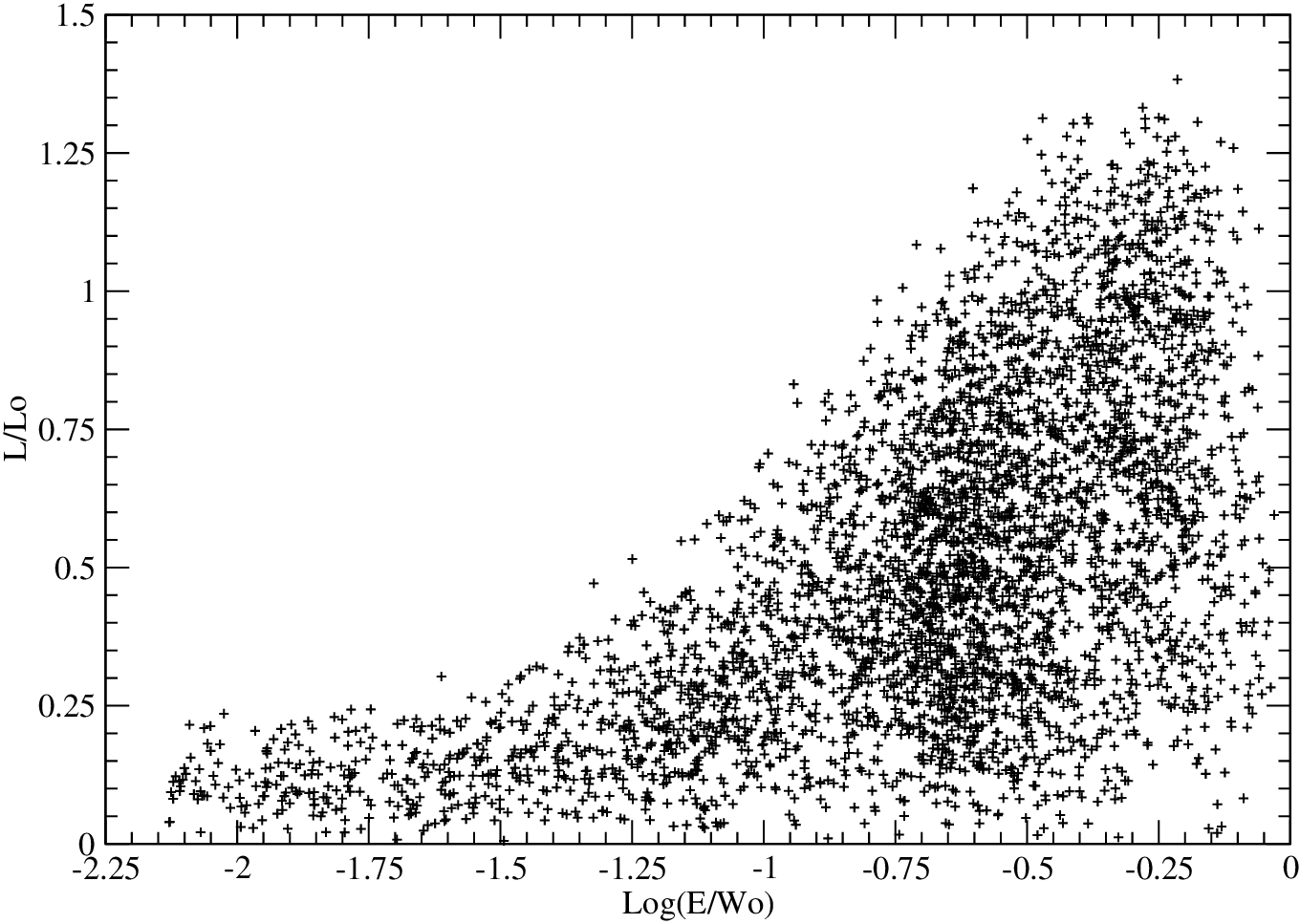}
\caption{Reduced energy versus reduced initial angular momentum for the original E4c model.
The upper and lower panels correspond, respectively, to regular and chaotic orbits.}
\label{E4corig}
\end{figure}

\begin{figure}
\vskip 11mm
\includegraphics[width=\hsize]{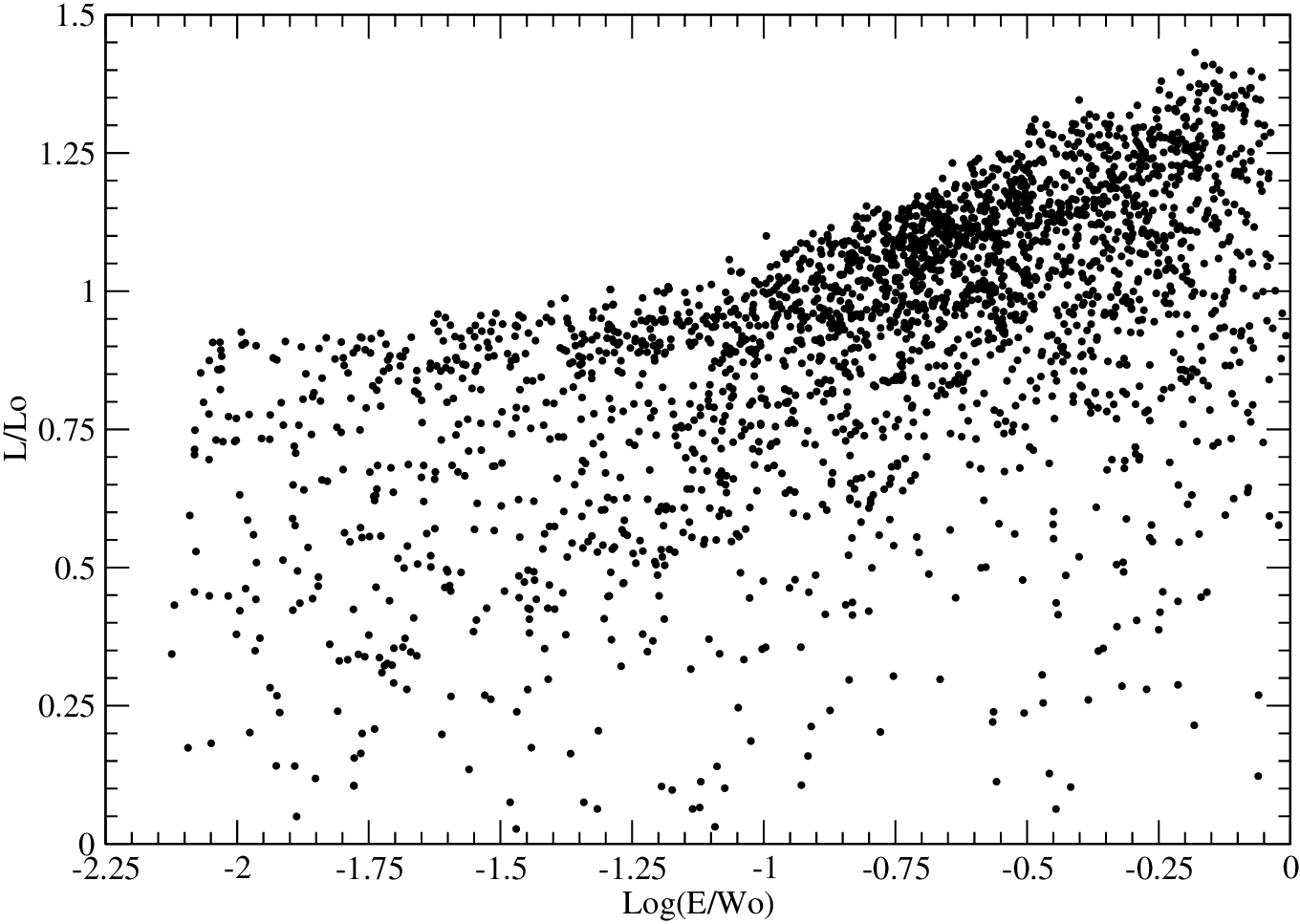}
\vskip 11mm
\includegraphics[width=\hsize]{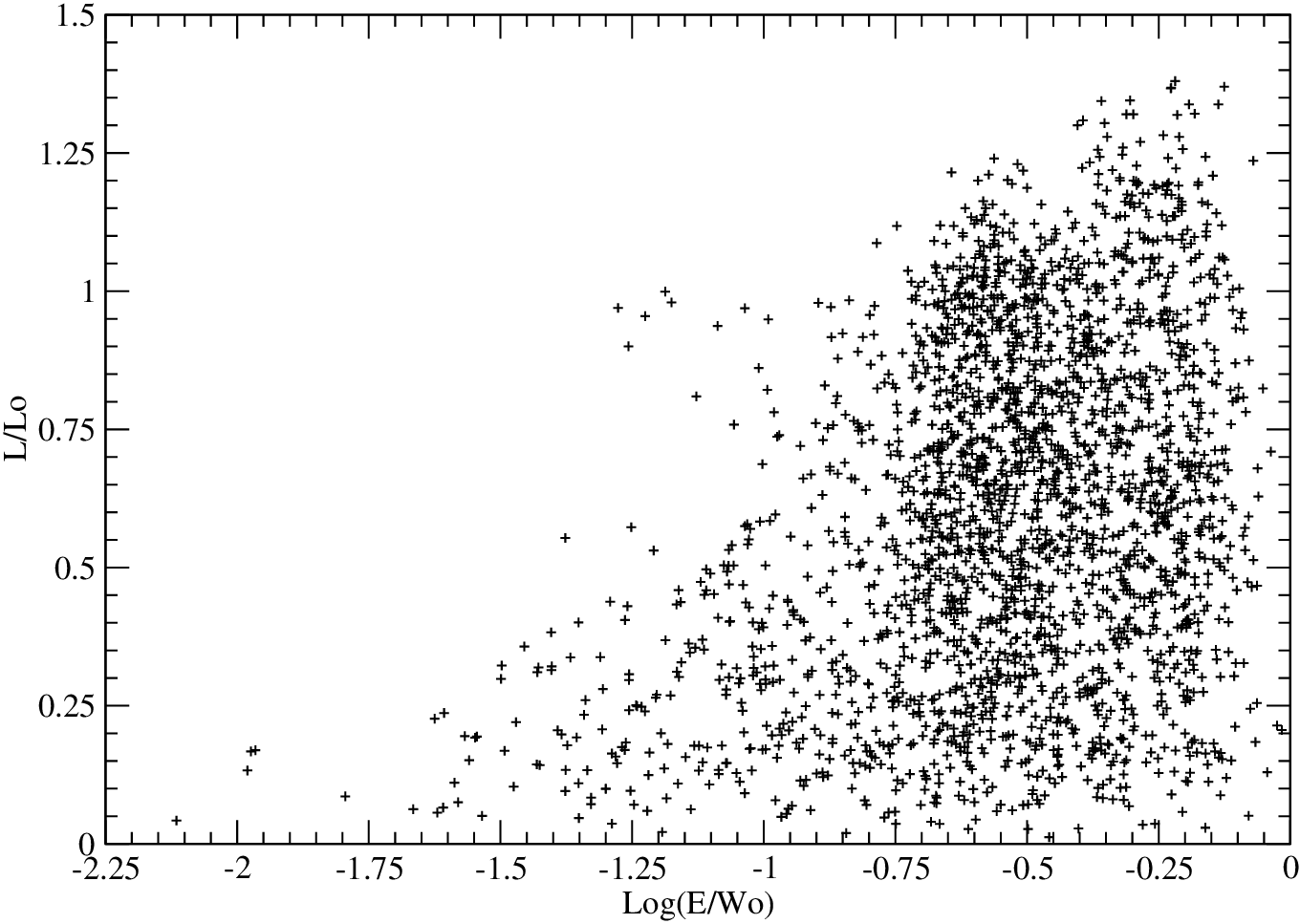}
\caption{Reduced energy versus reduced initial angular momentum for the isotropized E4ci model.
The upper and lower panels correspond, respectively, to regular and chaotic orbits.}
\label{E4isotr}
\end{figure}

Figs. \ref{E4corig} and \ref{E4isotr} present the reduced energy versus
reduced initial angular momentum, respectively, for the original, E4c, and the isotropized, E4ci,
models. It is quite clear the brutal cutoff to the larger angular momenta imposed by the
cold collapse used to create model E4c, and how it was compensated by our isotropization
technique in model E4ci. Regular and chaotic orbits (see below) are shown separately in
both Figures, and it is clear again that chaos is favored by low angular momenta. 

The price paid for the isotropization, in addition to the lower cusp slope,
was a rounder inertia tensor (see Table \ref{semiejes}). In order to try to
recover the original triaxiality, we followed the dragging recipe of \citet{HMSH01},
but we applied the squeezing in the $y$ (intermediate) direction only and we
modified the time parameters we had used in Section 2 to account for the
different values of $T_{\rm cr}$ in the two models. We tried several values of
$\xi_0$ between 0.5 and 0.8 and finally chose the models we had obtained with
$\xi_0=0.60$ and $\xi_0=0.80$, because they turned out to be the ones more similar
to models from \citet{ZM12}. After the squeezing had ended, we let the systems
relax for $100 T_{\rm cr}$ to obtain the final models, which we  call E4cf6
(the one obtained with $\xi_0=0.60$) and E4cf8 (corresponding to $\xi_0=0.80$).
We also evolved each one of these two models an additional $400 T_{\rm cr}$ interval
in order to check their stability and, again, all the changes of the macroscopic
parameters and $\gamma$ over a Hubble time were found to be of the order of one per cent.

Table \ref{semiejes} gives the axial ratios of the 20, 40, ..., 100 per cent most
tightly bound particles for the models. Following \citet{AM90}, the axial ratios of the
80 per cent most tightly bound particles are the ones we use to characterize the models,
and we see that those ratios are very similar for the present model E4cf6 and the E3 models
of \citet{ZM12}; also those ratios of our present model E4cf8 are not too different from
those of the E4 model of our previous paper. Unfortunately, for other fractions of most
tightly bound orbits the differences are larger and the present models tend to be less
flattened and less triaxial than those of the first paper in this series.
On the other hand, the anisotropy parameter $\beta$ shifted to values somewhat larger
than those of model E4ci, although they remain below 0.25 for most radii, and reach about
0.5 only in the  outermost regions (Fig. \ref{betae4c}). Thus, although the squeezing
process raised the anisotropy, we still ended with much more isotropic models
than the original E4c. The central cusp maintained a reasonably steep slope in both cases
($\gamma=0.924 \pm 0.024$ for the E4cf6 model and $\gamma=0.928 \pm 0.024$ for
the E4cf8 model).

\begin{table}
 \caption{Ratios of semiaxes corresponding to different percentages of most tightly
bound particles for our isotropized model and for that model after
the squeezing processes. The results for the E3b and E4c models of \citet{ZM12} are
also included for comparison.}
 \label{semiejes}
 \begin{tabular}{lcccccc}
  \hline
  Model & ratio & 20 & 40 & 60 & 80 & 100 \\
$ $ & $ $ &  per cent &  per cent &  per cent &  per cent & per cent\\
  \hline
  E4ci  & $b/a$ & 0.880 & 0.889 & 0.900 & 0.903 & 0.869 \\
        & $c/a$ & 0.705 & 0.685 & 0.693 & 0.715 & 0.800 \\
  E4cf8 & $b/a$ & 0.789 & 0.793 & 0.804 & 0.822 & 0.871  \\
        & $c/a$ & 0.668 & 0.663 & 0.655 & 0.615 & 0.790 \\
  E4cf6 & $b/a$ & 0.760 & 0.770 & 0.780 & 0.797 & 0.871 \\
        & $c/a$ & 0.712 & 0.714 & 0.714 & 0.683 & 0.786 \\
  E3b   & $b/a$ & 0.710 & 0.701 & 0.761 & 0.802 & 0.840 \\
        & $c/a$ & 0.554 & 0.555 & 0.634 & 0.692 & 0.803 \\
  E4c   & $b/a$ & 0.749 & 0.706 & 0.737 & 0.765 & 0.797  \\
        & $c/a$ & 0.588 & 0.515 & 0.539 & 0.575 & 0.708  \\
  \hline
 \end{tabular}
\end{table}

\subsection{Chaotic content}
Just as we had done for the model of Section 2, we used the LIAMAG routine
\citep{UP88} to obtain the FT-LCNs of 5,000 orbits of each one of the isotropized
models; due to the different time scales, the integration time was now
10,000 t.u. and the renormalization interval 1 t.u. Table
\ref{clae4cf} shows the results, together with those for models E3b and E4c
\citet{ZM12} for comparison, and the last column gives the triaxiality,
$T = (a^2 -b^2)/(a^2-c^2)$, evaluated from the 80 per cent most tightly bound
particles. Clearly, the percentages of regular orbits in our
isotropized models are much larger than those in the original collapse models,
with model E4ci having four or five times more regular orbits than the
collapse models of highest chaoticity. Unfortunately the structure of the
present models differs from the previous ones, so that we cannot be sure
that the difference is due to the different degrees of anisotropy, even though
it seems likely.

Reduced energy versus reduced initial angular momenta plots, separating
the regular and chaotic orbits, were also done for the E4cf8 and E4cf6 models but,
as they turned out to be similar to the one for the E4ci model, of Fig. \ref{E4isotr},
they are not shown.

\begin{table}
 \caption{Percentages of regular, partially chaotic and fully chaotic orbits in
the isotropized models. The results for the E3b and E4c models of \citet{ZM12},
and the triaxiality of each model, $T$, are also included for comparison.}
 \label{clae4cf}
 \begin{tabular}{llccccccc}
  \hline
  Model & Regular & Part. Chaotic & Fully Chaotic & $T$ \\
$ $ & (per cent) & (per cent) &  (per cent) & \\
  \hline
E4ci & $50.16\pm0.71$ & $9.18\pm0.41$ & $40.66\pm0.69$ & 0.38\\
E4cf8 &  $41.28\pm0.70 $ & $9.24\pm0.41$  & $49.48\pm0.71$ & 0.52 \\
E4cf6 & $45.88\pm0.70$ & $11.04\pm0.44$ & $43.08\pm0.70$ & 0.68\\
E3b   &  $14.04\pm0.50$ & $13.57\pm0.49$ & $72.39\pm0.64$ & 0.68 \\
E4c   &  $12.67\pm0.49$ & $10.11\pm0.45$ & $77.22\pm0.62$ & 0.62 \\
  \hline
 \end{tabular}
\end{table}

\begin{table}
\caption{Orbital classification of the regular orbits in the isotropized
models. The results for the E3b and E4c models of \citet{MNZ13} are also
included for comparison.}
\label{regul}
\begin{tabular}{@{}lcccc@{}}
\hline
Model & BBL & SAT & ILAT & OLAT \\
$ $ &  (per cent) & (per cent) & (per cent) & (per cent) \\ 
E4ci & 1.72 $\pm$ 0.26 & 81.53 $\pm$ 0.78 & 0.16 $\pm$ 0.08 & 16.59 $\pm$ 0.74 \\ 
E4cf8 & 4.43 $\pm$ 0.45 & 73.98 $\pm$ 0.97 & 1.07 $\pm$ 0.23 & 20.52 $\pm$ 0.89 \\ 
E4cf6 & 3.23 $\pm$ 0.37 & 59.02 $\pm$ 1.03 & 4.40 $\pm$ 0.43 & 33.35 $\pm$ 0.98 \\
E3b & 23.00 $\pm$ 1.61 & 53.86 $\pm$ 1.90 & 6.84 $\pm$ 0.96 & 13.10 $\pm$ 1.29 \\ 
E4c & 26.78 $\pm$ 1.85 & 66.43 $\pm$ 1.97 & 4.00 $\pm$ 0.82 & 0.70 $\pm$ 0.35 \\ 
\hline
\end{tabular}
\end{table}

\section{Regular orbits}
As in our previous papers \citep*{AMNZ07,MNZ09,MNZ13}, we used
frequency analysis to classify the regular orbits and the results are presented
in Table \ref{regul} together with those for our E3b and E4c previous models,
for comparison. There are clear decreases in the percentages of boxes and boxlets
(BBLs) and clear increases in the percentages of outer long axis tubes (OLATs) as
we go from the original to the isotropized models. The fraction of small axis tubes
(SATs) seems to be larger for the E4ci and E4cf8 models, but these models have smaller
triaxility than models E3b and E4c (see Table \ref{clae4cf}); model E4cf6, whose
triaxiality is equal to that of model E3b, and not too different from that of model
E4c, has a fraction of SATs similar to the original models. The percentages of inner
long axis tubes (ILATs) seem to be smaller in the isotropized models, but these
fractions are generally low anyway.

\section{Conclusions}
We have used the adiabatic dragging of \citet{HMSH01} to build up a cuspy
triaxial model similar to the one they had investigated. Using a longer
relaxation time than the one they had used we found that the model had
not reached equilibrium at the latter time. The cusp, in particular, steadily
reduces its slope and, after $100 T_{\rm cr}$, it is only $\gamma \simeq 0.8$.
Nevertheless, we searched for chaotic orbits in that model as \citet{HMSH01} had
done and we found that, on the one hand, the frequency difference technique
they had used is not adequate and is the likely culprit of their finding
almost no chaos and, on the other hand, that when other techniques
(such as Lyapunov exponents and SALI) are used, it turns out that more than one
fourth of the orbits are chaotic. Moreover, chaoticity affects the distribution
of orbits, so that it should be taken into account in the study of the
structure of the model. The model of \citet{HMSH01} was the only cuspy and triaxial
system built with the $N$--body method where almost no chaos had been found, so that
our result solves that puzzle.

Nevertheless, the chaotic fraction we found is still substantially lower than that
of other models built with the $N$--body method \citep{MNZ09, ZM12}, so that we
investigated whether the extreme velocity anisotropy of the latter could explain
that difference. We succeeded in rendering the velocity distribution of one of the
models of \citet{ZM12} much less radially oriented and, thus, more isotropic and
we also applied the adiabatic drag to the isotropized model to obtain two other
more triaxial systems. The chaotic fractions in these models are in the 50 to 60
percent range, substantially lower than the 75 to 90 per cent range found in the
models of  \citet{ZM12}. Unfortunately, although the semiaxial ratios of the
80 per cent most tightly bound particles of the isotropized models are similar
to those of some of the original models, those ratios differ for other fractions
of most tightly bound particles, so that the models are not strictly comparable.
Nevertheless, the fact that $\it all$ the isotropized models have lower chaoticity
than $\it any$ of the original collapse models strongly suggests that the radially
oriented orbits of the latter favored the onset of chaoticity.

Finally, our results clearly show that one can obtain cuspy triaxial models with
fairly isotropic velocity distributions that contain large fractions of chaotic
orbtis and, nevertheless, are highly stable so that, in this sense, they extend our
similar previous results for models with strongly radial velocity distributions.

\section*{Acknowledgments}

We are very grateful to L. Hernquist, D. Nesvorn\'y 
and D. Pfenniger for allowing us to use their
codes, to R.E. Mart\'{\i}nez and H.R Viturro for their technical assistance and
to A.F. Zorzi for her help. This work was supported with grants from the
Consejo Nacional de Investigaciones Cient\'{\i}ficas y T\'ecnicas de la
Rep\'ublica Argentina, the Agencia Nacional de Promoci\'on Cient\'{\i}fica y
Tecnol\'ogica, the Universidad Nacional de La Plata and the Universidad Nacional
de Rosario.

\bsp

\label{lastpage}

\end{document}